# Selective synthesis of large-area monolayer tin sulfide from simple substances

Kazuki Koyama, Jun Ishihara, Takeshi Odagawa, Makito Aoyama, Chaoliang Zhang, Shiro Entani, Ye Fan, Atsuhiko Mori, Ibuki Kitakami, Sota Yamamoto, Toshihiro Omori, Yasuo Cho, Stephan Hofmann, and Makoto Kohda[*]

**Corresponding Author**

**Makoto Kohda** – *Department of Materials Science, Graduate School of Engineering, Tohoku University, Sendai, Japan; Center for Science and Innovation in Spintronics (Core Research Cluster), Tohoku University, Sendai, Japan; Division for the Establishment of Frontier Science, Tohoku University, Sendai, Japan; Quantum Materials and Applications Research Center, National Institute for Quantum Science and Technology, Gunma, Japan;* [*]E-mail: makoto.koda.c5@tohoku.ac.jp

**Authors**

**Kazuki Koyama** – *Department of Materials Science, Graduate School of Engineering, Tohoku University, Sendai, Japan*

**Jun Ishihara** – *Department of Materials Science, Graduate School of Engineering, Tohoku University, Sendai, Japan*

**Takeshi Odagawa** – *Department of Materials Science, Graduate School of Engineering, Tohoku University, Sendai, Japan*

**Makito Aoyama** – *Department of Materials Science, Graduate School of Engineering, Tohoku University, Sendai, Japan*

**Chaoliang Zhang** – *Department of Applied Physics, Graduate School of Engineering, Tohoku University, Sendai, Japan*

**Shiro Entani** –*, National Institutes for Quantum Science and Technology, Gunma, Japan*

**Ye Fan** – *Electrical Engineering Division, Department of Engineering, University of Cambridge, Cambridge, United Kingdom*

**Atsuhiko Mori** – *Department of Materials Science, Graduate School of Engineering, Tohoku University, Sendai, Japan*

**Ibuki Kitakami** – *Department of Materials Science, Graduate School of Engineering, Tohoku University, Sendai, Japan*

**Sota Yamamoto** – *Department of Materials Science, Graduate School of Engineering, Tohoku University, Sendai, Japan*

**Toshihiro Omori** – *Department of Materials Science, Graduate School of Engineering, Tohoku University, Sendai, Japan*

**Yasuo Cho** – *New Industry Creation Hatchery Center, Tohoku University, Sendai, Japan*

**Stephan Hofmann** – *Electrical Engineering Division, Department of Engineering, University of Cambridge, Cambridge, United Kingdom*

**Abstract**

Both tin monosulfide (SnS) and tin disulfide ($SnS_2$) are thermodynamically stable layered materials with potential for spin–valleytronic devices and photodetectors. Notably, monolayer SnS, owing to its low symmetry, exhibits interesting properties such as ferroelectricity, shift-current, and a persistent spin helix state in the monolayer limit. However, creating atomic-thickness crystals of SnS is challenging owing to the enhanced interlayer interactions caused by lone pair electrons, unlike to $SnS_2$. Here, we demonstrate that p-type SnS can be selectively grown by simply varying the sulfur vapor concentration relative to tin using single-element precursors. We show that monolayer SnS crystals, up to several tens of micrometers in lateral scale, can be easily and safely obtained by high-temperature etching of bulk SnS in a pure nitrogen gas atmosphere. These findings pave the way for device applications based on high-quality tin sulfide.

**Main text**

Tin sulfide systems, including chemically stable layered semiconductors such as tin monosulfide (SnS) and tin disulfide ($SnS_2$), have attracted considerable attention owing to their abundance, cost-efficiency, and excellent electrical and optical properties for valleytronics and spintronics. Bulk SnS, which are part of the *Pnma* space group at room temperature and ambient pressure, exhibits a puckered structure along the armchair direction with high in-plane anisotropy between the zigzag and armchair directions, as shown in Fig. 1a. Unlike transition metal dichalcogenides (TMDs), which possess energetically degenerate non-equivalent valleys[1], SnS has two sets of valleys with different bandgaps[2,3]. This opens new possibilities for valleytronics by enabling a clear distinction of the valley from the emitted photon energy. When thinned to a monolayer, SnS belongs to the *Pmn*$2_1$ space group. Symmetry lowering leads to the lack of spatial inversion symmetry and results in the emergence of pure in-plane ferroelectricity at room temperature[4], and the shift-current photovoltaics[5] have been experimentally demonstrated. A polarizer of emitting only X- or Y-polarized light by electrically switching the direction of ferroelectric polarization has been proposed[6]. Furthermore, in p-type monolayer SnS, a large uniaxial spin-orbit field acting on holes is theoretically predicted in the out-of-plane direction, attributed to its point group symmetry and the heavy constituent atoms of SnS[7]. This is known as the persistent spin helix state, which is expected to enable the development of spin logic devices through coherent spin propagation[8–10] protected by SU(2) symmetry[11]. As a result, monolayer SnS is a promising candidate for novel spin/valley-control functions, such as spin texture inversion through ferroelectric polarization switching[7].

SnS has been selectively grown by the chemical vapor deposition (CVD) method using SnCl$_4$ and H$_2$S[12], SnO$_2$ and S[13], and SnS and S[14], or by the chemical vapor transport method using I$_2$ as a carrier agent in addition to Sn and S[15]. However, when crystals are grown from precursors containing elements other than Sn and S, there remains a risk of impurity element contamination in the grown crystals, similar to what is observed in TMDs[16,17]. Moreover, high-purity SnS powder used as a precursor is relatively expensive. Therefore, it is preferable to grow SnS crystals using only the single elements Sn and S as precursors. Additionally, the spread of lone electron pairs into the interlayer spaces enhances the interlayer interactions in SnS[18]. This makes synthesizing large-area monolayer SnS difficult. This characteristic contrasts with SnS$_2$, where monolayer SnS$_2$ can be obtained through mechanical exfoliation[19] because the atomic layers are stacked by pure van der Waals forces in the hexagonal $P\bar{3}m1$ space group (Fig. 1b).

Here, we selectively grow pure bulk SnS from the simplest precursors, Sn and S, using the CVD method, and then convert bulk crystals to monolayers by etching. By controlling the relative concentration of S precursor vapor to Sn precursor vapor during the chemical reaction, pure SnS can be grown over large areas exceeding 14,000 μm$^2$. Adjusting the S concentration by varying the distance between the precursors allows for the selective growth of SnS$_2$, mixtures of SnS and SnS$_2$, as well as SnS. Furthermore, by etching the grown bulk SnS at high temperatures in a pure nitrogen atmosphere, mono-to-few-layer SnS crystals, several tens of micrometers in size, are obtained. This method is simpler and safer for synthesizing large-area monolayer SnS compared to previous methods that required stringent growth conditions[4] and highly toxic source materials[20]. This study paves the way for the development of novel optical and spin–valley devices based on high-quality tin sulfides.

To explore selective growth methods of SnS from single-element precursors, we calculate the S–Sn binary phase diagram including the gas phase at $10^5$ Pa, using the thermodynamic model and parameters of the S–Sn system[21] by Thermo-Calc Software. The stability regions of each tin sulfide can be categorized into three cases based on the atomic percentage of sulfur, as depicted in Fig. 1c.

(i) 0 < at% S < 50: Only $\alpha$(*Pnma*)/$\beta$(*Cmcm*)-SnS is stable. The liquid phase, formed during high-temperature reaction mixing, disappears upon cooling, triggering a phase transition from $\beta$-SnS to $\alpha$-SnS below approximately 600°C[22], which should result in $\alpha$-SnS at room temperature.

(ii) 50 < at% S < 67: SnS and $Sn_2S_3$ are stable at 50 < at% S < 60, and $SnS_2$ and $Sn_2S_3$ are stable at 60 < at% S < 67. Because produced $Sn_2S_3$ is likely to be thermodynamically metastable[23,24], it may decompose to stable SnS and $SnS_2$[18].

(iii) 67 < at% S < 100: Only $SnS_2$ is stable. The liquid and gas phases, which are mixed during the reaction at high temperatures, should disappear upon cooling.

Based on this phase diagram, SnS and $SnS_2$ can be selectively grown from single-element precursors using a straightforward chemical reaction method in regions of low (high) sulfur concentration where the SnS ($SnS_2$) phase is stable.

SnS and SnS$_2$ crystals are grown using the CVD method in a horizontal quartz tube furnace with two heating zones, as illustrated in Fig. 1d. The precursors used are sulfur powder (99.99%, Kojundo Chemical Laboratory Co., Ltd.) and tin powder (99.99%, Kojundo Chemical Laboratory Co., Ltd.). Ceramic boats containing 0.7 g of sulfur and 0.3 g of tin are placed upstream and downstream of the quartz tube, respectively. A SiO$_2$/Si substrate is positioned on top of the downstream boat, with the growth surface facing downward. The quartz tube is purged with Ar gas three times before the process. Crystal growth is performed under Ar gas at a flow rate of 100 sccm. The heater for the sulfur powder is ramped up to 200°C over 30 min and maintained for another 30 min, while the heater for the tin powder and substrate is ramped up to 850°C over 30 min and held for 30 min. After the heating process, both boats are allowed to cool naturally. Because some of the evaporated sulfur gas solidifies in the quartz tube section between the two heaters, exposed to room temperature, the amount of sulfur reaching the boat containing tin and substrate varies with the distance $d$ between the heaters. Therefore, adjusting $d$ allows accurate control of the sulfur concentration, facilitating selective growth based on the phase diagram.

Figure 1e depicts microscope images of crystals obtained with varying distances $d$ between the two heaters. Crystals with a rhombic shape are observed at $d = 9.5$ cm, indicating the presence of orthorhombic SnS[25,26]. When the distance is decreased to $d = 4.5$ cm (i.e., sulfur concentration increases), crystals with hexagonal facets are observed, indicating the growth of SnS$_2$. At an intermediate distance of $d = 8.0$ cm, orthorhombic-shaped structures are visible, but some surfaces appear rough, suggesting a mixture of different phases in the crystal. These findings demonstrate the selective growth of different crystal structures by adjusting the distance $d$. To achieve large-area SnS exceeding 10,000 µm$^2$, positioning the substrate downstream from the Sn boat is effective. In this configuration, the molecular vapor pressure on the substrate is lower compared to a closed environment around the substrate[27]; in this setup, the substrate is positioned facing downward on the boat, as depicted in Fig. 1d. Consequently, the nucleation density in the growth zone on the substrate decreases, promoting the spread and growth of crystals from individual nuclei. Figure 1f displays microscope images of SnS crystals grown under these conditions, where the substrate is placed downstream from the boat with the growth surface facing upward, and the holding time at the maximum temperature of both heaters is extended to 60 min. Large crystals exceeding 150 µm in size are obtained using this method. The low molecular vapor pressure is expected to facilitate selective SnS growth. According to the S–Sn phase diagram calculated at 0.1 Pa (as shown in Fig. S1 in the supporting information), at temperatures ≳700 K, the stability of SnS$_2$ and Sn$_2$S$_3$ decreases across all sulfur concentrations, favoring the stability of SnS structures instead. By suppressing the formation of compounds with stoichiometric ratios exceeding Sn:S = 1:1, phase-pure SnS can be obtained through rapid cooling from high growth temperatures without needing to adjust the sulfur concentration, thereby enabling large-area growth of SnS.

To identify the phase of the three crystals in Fig. 1e, angle-resolved Raman spectroscopy is performed. As illustrated in Fig. 2a, the crystal grown at $d$ = 9.5 cm exhibits four distinct Raman peaks[13], A$_g$(1) (94 cm$^{-1}$), A$_g$(2) (189 cm$^{-1}$), A$_g$(3) (216 cm$^{-1}$), and B$_{3g}$ (163 cm$^{-1}$), indicating pure SnS. For $d$ = 4.5 cm, as shown in Figure 2c, the crystal grown exhibits a very weak $E_g$ mode (206 cm$^{-1}$) and strong $A_{1g}$ mode (315 cm$^{-1}$), confirming pure SnS$_2$[13]. In the intermediate condition at $d$ = 8.0 cm, four Raman peaks specific to SnS are observed in the smooth surface region. However, in the rough surface region, an additional $A_{1g}$ peak specific to SnS$_2$ is also observed. This indicates a crystal mixture of SnS and SnS$_2$. Figures 2d and 2e depict the angle dependence of the Raman peak intensities for SnS and SnS$_2$, obtained by rotating the crystals relative to a linearly polarized excitation laser. For SnS, the $B_{3g}$ mode exhibits fourfold symmetry, while the $A_g$(1)–(3) modes display twofold symmetry, as shown in Fig. 2d. This anisotropy reveals that the armchair (zigzag) direction aligns with the bisector of the acute (obtuse) angle of SnS grown at thermodynamic equilibrium, consistent with a previous study[28]. Conversely, the nearly constant $A_{1g}$ peak intensity from SnS$_2$, shown in Fig. 2e and independent of the angle, indicates that SnS$_2$ possesses an in-plane isotropic structure.

The microscopic crystal structures of SnS and SnS$_2$ are assessed using high-angle annular dark field STEM images obtained with a scanning transmission electron microscope (STEM) (Titan$^3$ 60-300 Double Corrector, Thermo Fisher Scientific). Figures 3a and 3b display the cross-sectional and top-view (c-plane) STEM images of grown SnS and SnS$_2$, where the positions of Sn and S atoms are identified by energy dispersive X-ray spectroscopy (EDS) mapping. However, the top-view elemental mapping of SnS does not show distinct Sn and S atoms owing to their overlapping positions. In Fig. 3a, a puckered structure of each SnS layer along the armchair direction is clearly observed in the cross-sectional STEM image, and a rectangular unit cell is observed in the top-view image. The bulk SnS obtained is identified as the *α*-phase, characterized by its puckered cross-section and the stacking of atomic sheets in a phase-reversed manner[29]. This indicates that SnS grown on SiO$_2$/Si substrates above the phase transition temperature can change to the *α* phase during cooling[14] without the *β* phase[30] being frozen. For SnS$_2$, the cross-sectional STEM image reveals a structure where an Sn atomic sheet is sandwiched between two S atomic sheets, while the top-view image displays a hexagonal unit cell, as depicted in Fig. 3b. The well-ordered atomic arrangement in these STEM images confirms the high crystallinity of the grown SnS and SnS$_2$.

To quantitatively analyze the Sn and S compositions in the SnS and SnS$_2$ crystals, we perform energy dispersive X-ray spectroscopy (EDS) using a scanning electron microscope. Figures 3c and 3d present the EDS spectra of SnS and SnS$_2$. Both spectra show peaks originating only from Sn and S, alongside minor peaks attributed to Si elements from the SiO$_2$/Si substrate. No elements besides Sn and S are detected above the EDS detection limit of approximately 0.1 at%, except for elements originating from the substrate. This confirms the growth of high-purity SnS and SnS$_2$ using single-element Sn and S precursors. Additionally, we perform electron backscatter diffraction (EBSD) measurements to determine the crystal orientations. Figures 3e and 3f depict EBSD patterns and inverse pole figure maps for SnS and SnS$_2$ crystals. The distinct EBSD patterns of both crystals indicate their high crystallinity, while the inverse pole figure maps for the normal direction reveal that the entire surface of SnS is oriented along (001), and that of SnS$_2$ is oriented along (0001), indicating *c*-plane-oriented single crystals. Further structural analyses confirmed the growth of large-area, single-phase SnS crystals. The electronic structure analysis also identifies the SnS and SnS$_2$ phases. Figures 3g and 3h show fluorescent-yield S *K*-edge X-ray absorption near edge structure (XANES) spectra of SnS and SnS$_2$ measured at BL-08W of NanoTerasu (Sendai, Japan). The S *K*-edge XANES technique is based on X-ray absorption by S atoms, thereby providing information on the SnS and SnS$_2$ crystals that is fee from the influence of the substrate. The results obtained for SnS and SnS$_2$ show significant differences for the electronic states in the vicinity of the absorption edge. A broad peak is observed at about 2471 eV for SnS. In contrast, three sharp peaks are observed at 2496.1 eV, 2472.8 eV and 2477.2 eV for SnS$_2$. These features of the XANES spectra are in accordance with previous reports of SnS and SnS$_2$ and are well reproduced by *ab initio* XANES simulations with FDMNES[31–33]. The XANES results obtained here for SnS and SnS$_2$ exhibit evidence for the growth of single-phase SnS and SnS$_2$ crystals, respectively.

The p-type SnS is required to utilize the long spin coherence through the persistent spin helix state[7]. The carrier types in the grown SnS are investigated using Scanning Nonlinear Dielectric Microscopy (SNDM)[34,35], which can directly assess localized carrier type information without the need for electrode fabrication, even for atomically thin layered materials[36,37]. Figure 4 displays the topographic and voltage-dependent differential capacitance d*C*/d*V* images of the SnS crystals. The positive d*C*/d*V* signal in regions corresponding to the topographic contrast indicates that the grown SnS is p-type. This observation aligns with previous findings attributing the p-type nature of SnS to Sn vacancies[13,15,38].

Bulk SnS is etched in a high-temperature nitrogen atmosphere to obtain large-area monolayer crystals of SnS. The enhanced growth of SnS in the out-of-plane direction owing to strong interlayer interactions[39] can be countered by thinning after the growth of large-area bulk crystals. The strong bonding between the substrate and the bottom of the crystals prevents the etching of the lowest layered structure of SnS[40]. The SiO$_2$/Si substrate, upon which bulk SnS is grown, is positioned in the same quartz tube as depicted in Fig. 1d, with the side of the crystals facing upward. The temperature is ramped up to 700°C over 20 min under a nitrogen gas flow rate of 100 sccm, held for 20 min, and then allowed to cool naturally. The crystal thickness is determined using tapping mode atomic force microscopy (AFM) (Bruker Dimension Icon, Asylum MFP-3D AFM). Figures 5a and 5b present optical microscopy and AFM images of the crystals after etching, respectively, with the height profile along the white solid line in Fig. 5b shown in Fig. 5c. The crystal thickness is determined to be 0.87 nm, consistent with the previously reported thickness of approximately 0.8 nm for monolayer SnS[4]. Some crystals obtained under this etching condition measure approximately 1.8 nm in thickness, corresponding to 2 or 3 layers, as depicted in Fig. 5d. Even under a relatively "gentle" etching condition of maintaining 700°C for 10 min with a nitrogen flow of 30 sccm, crystals as thin as 1.5 nm (bilayer) are obtained, as shown in Fig. 5e. This nitrogen etching method facilitates the synthesis not only of monolayer[40] but also bilayer SnS, wherein the application of a gate voltage induces breaking of inversion symmetry owing to the electric field[41] by tunning heating time and a flow rate of etching gas.

In conclusion, we have successfully achieved selective growth of SnS through CVD using only the single elements Sn and S. Adjusting the S precursor concentration relative to the Sn precursor concentration allows for the growth of SnS at lower S concentrations and $SnS_2$ at higher S concentrations. Even when SnS is grown at high temperatures where the high-symmetry nonpolar $β$ phase is stable, natural cooling induces a phase transition to the low symmetry polar $α$ phase. By applying a high-temperature etching method using pure nitrogen gas to SnS, large-area monolayers and few-layer SnS can be easily and safely synthesized. These results offer new insights into the development of novel applications using ultrathin tin sulfides.

**Associated content**

**Supporting information**

Phase diagram of the S–Sn system calculated at 0.1 Pa

**Acknowledgments**

The authors thank Dr. Takamichi Miyazaki, Yuichiro Hayasaka, and Professor Daisuke Tadaki at Tohoku University for providing technical assistance. K.K and T.O acknowledge financial support from the Graduate Program in Spintronics (GP-Spin) at Tohoku University. This research was supported by JSPS KAKENHI Grant No. 21H04647, JST

FOREST and CREST programs (Grant Nos. JPMJFR203C and JPMJCR22C2). S.H. acknowledges funding from EPSRC (EP/P005152/1).

(41) Bao, Y.; Song, P.; Liu, Y.; Chen, Z.; Zhu, M.; Abdelwahab, I.; Su, J.; Fu, W.; Chi, X.; Yu, W.; Liu, W. Gate-Tunable In-Plane Ferroelectricity in Few-Layer SnS. *Nano Lett.* **2019**, *19* (8), 5109–5117. https://doi.org/10.1021/acs.nanolett.9b01419.

**Figure 1**

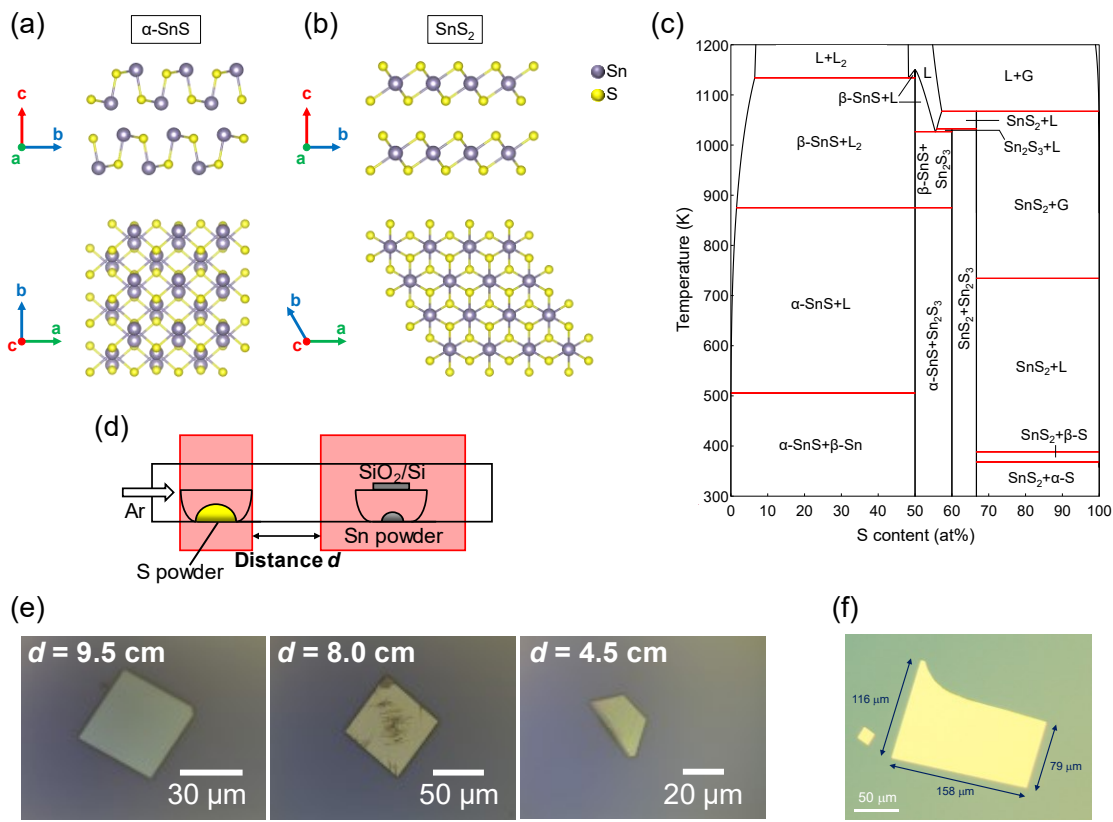

(a) Illustration of the crystal structure of α-SnS, stable at room temperature and ambient pressure. (b) Illustration of the crystal structure of SnS$_2$. (c) The phase diagram of the S–Sn system calculated at 10$^5$ Pa, where L represents the liquid phase and G the gas phase. (d) Schematic of the CVD system used for growing SnS and SnS$_2$ crystals. (e) Optical images of crystals obtained under conditions of $d$ = 9.5, 8.0, and 4.5 cm. (f) Optical image of a large-area bulk SnS crystal

**Figure 2**

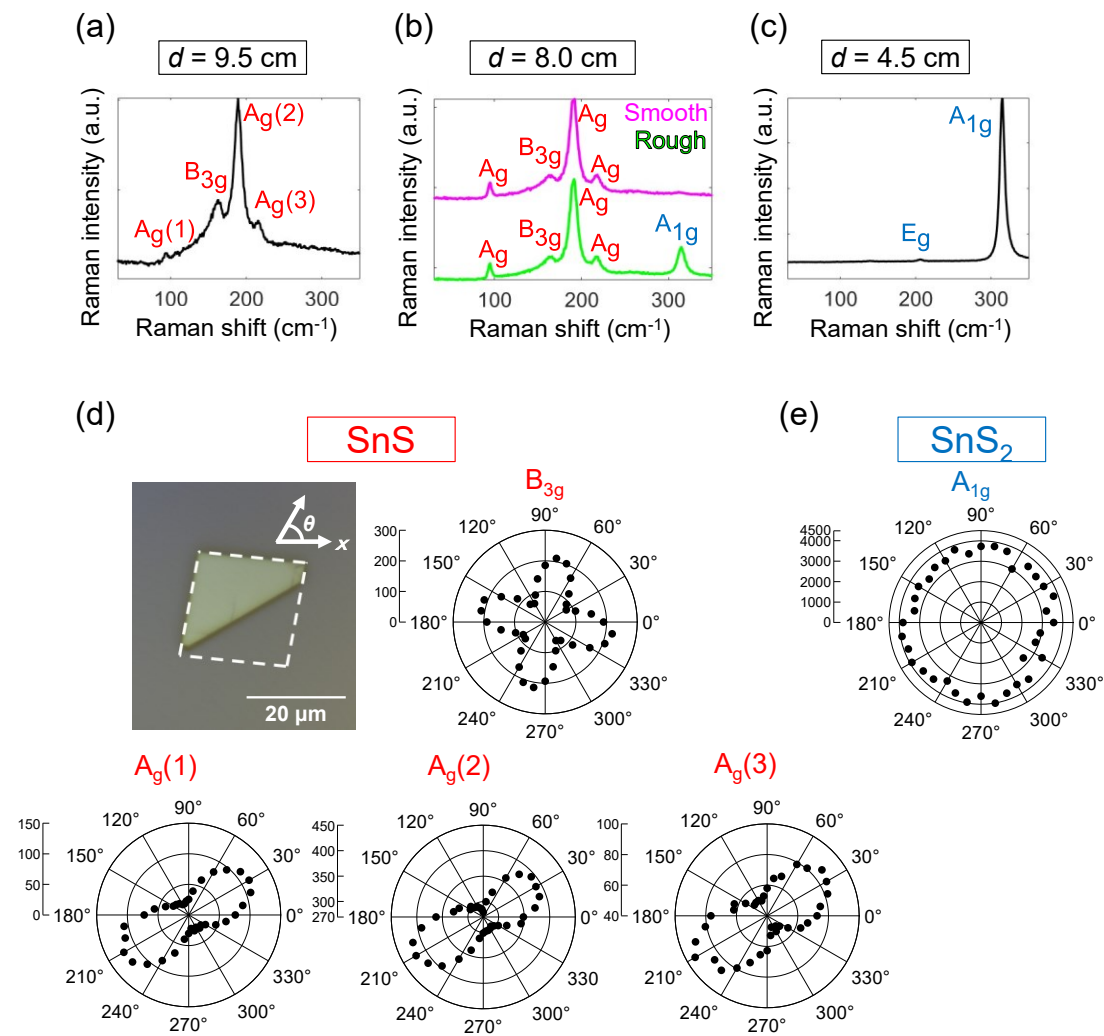

Raman spectra from the crystals grown under conditions (a) $d = 9.5$ cm, (b) $d = 8.0$ cm, and (c) $d = 4.5$ cm. (d) Crystal rotation angle dependence of Raman intensity for $B_{3g}$, $A_g(1)$, $A_g(2)$, and $A_g(3)$ modes in the Raman spectra of SnS crystals. (e) Crystal rotation angle dependence of $A_{1g}$ Raman peak intensity in $SnS_2$ crystals

**Figure 3**

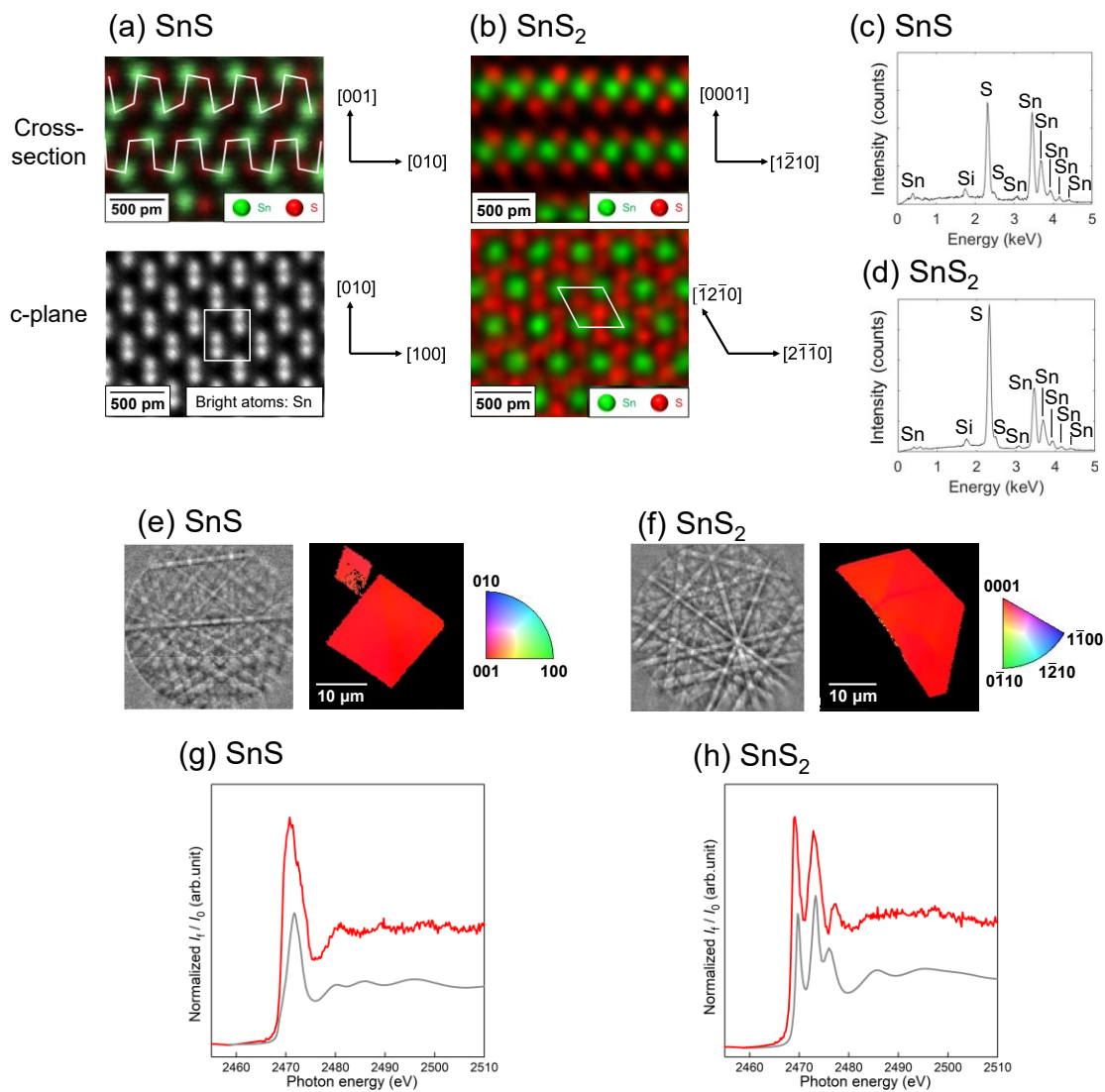

(a) Cross-sectional and top-view STEM images of the SnS crystal. Atomic positions of Sn and S in the cross-sectional image are identified by EDS mapping. (b) STEM images of the SnS$_2$ crystal with EDS mapping. The white solid line represents the unit cell. SEM–EDS spectra of (c) SnS and (d) SnS$_2$. Calculated atomic ratios for each element are also provided. SEM–EBSD patterns (left) and inverse pole figure maps for normal direction (right) of (e) SnS and (f) SnS$_2$ crystals. S *K*-edge XANES spectra of (g) SnS and (h) SnS$_2$ (red lines). The calculated XANES spectra from the FDMNES simulation are also shown in the same figure (gray lines)

**Figure 4**

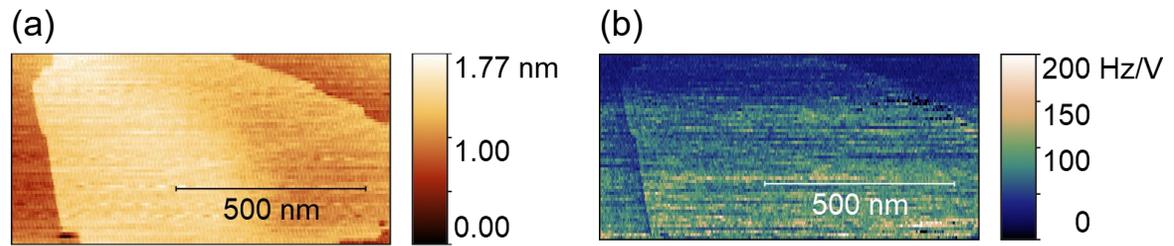

(a) Topographic and (b) d*C*/d*V* images of grown SnS crystal

**Figure 5**

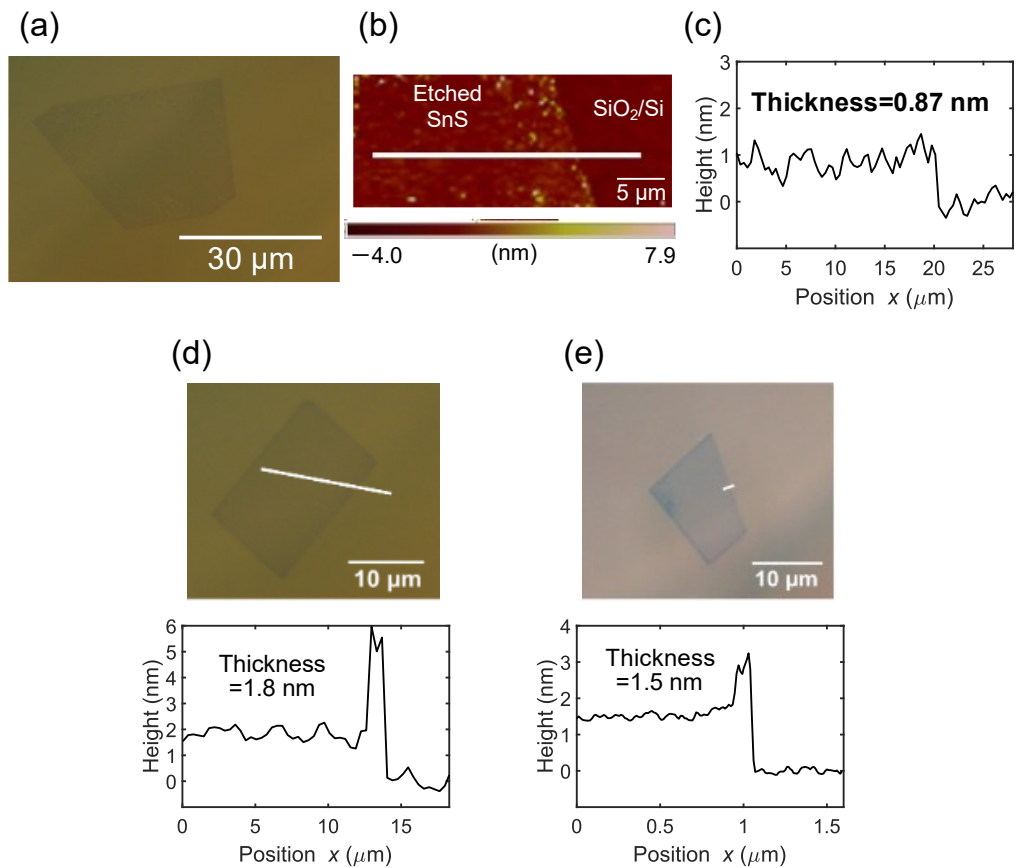

(a) Optical image of the SnS thin film remaining after etching bulk SnS on a SiO$_2$/Si substrate. (b) AFM image of the crystal shown in (a). (c) AFM height profile along the white solid line in (b). (d) SnS thin film corresponding to 2–3 layers obtained under the same etching conditions as (a). (e) SnS thin film corresponding to a bilayer left after etching at 700°C for 10 min under 30 sccm nitrogen flow

## Selective synthesis of large-area monolayer tin sulfide from simple substances

Kazuki Koyama[1], Jun Ishihara[1], Takeshi Odagawa[1], Makito Aoyama[1], Chaoliang Zhang[2], Shiro Entani[3], Ye Fan[4], Atsuhiko Mori[1], Ibuki Kitakami[1], Sota Yamamoto[1], Toshihiro Omori[1], Yasuo Cho[5], Stephan Hofmann[4], and Makoto Kohda[1,6-8]*

[1]*Department of Materials Science, Graduate School of Engineering, Tohoku University, Sendai, Japan*

[2]*Department of Applied Physics, Graduate School of Engineering, Tohoku University, Sendai, Japan*

[3]*National Institutes for Quantum Science and Technology, Gunma, Japan*

[4]*Electrical Engineering Division, Department of Engineering, University of Cambridge, Cambridge, United Kingdom*

[5]*New Industry Creation Hatchery Center, Tohoku University, Sendai, Japan*

[6]*Center for Science and Innovation in Spintronics (Core Research Cluster), Tohoku University, Sendai, Japan*

[7]*Division for the Establishment of Frontier Science, Tohoku University, Sendai, Japan*

[8]*Quantum Materials and Applications Research Center, National Institute for Quantum Science and Technology, Gunma, Japan*

*E-mail: makoto.koda.c5@tohoku.ac.jp

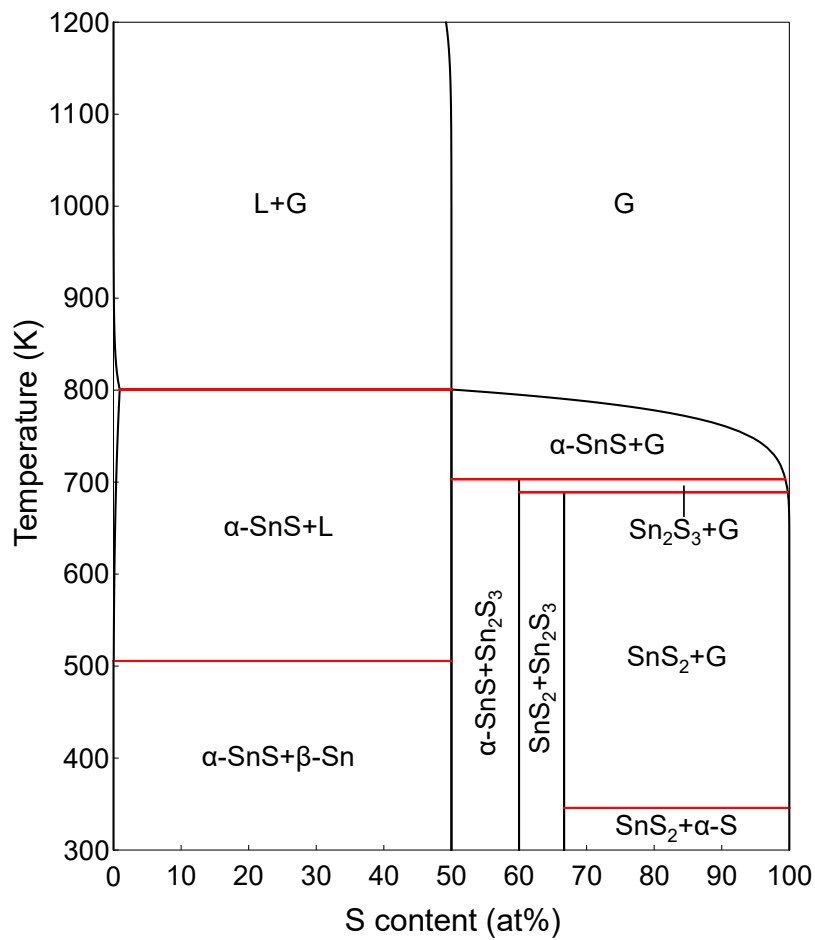

**Figure S1. Phase diagram of the S–Sn system calculated at 0.1 Pa.** The tdb file by Guan et al.[1] is used.

**Supplementary Reference**